\documentclass[12pt]{iopart}
\usepackage{graphicx}

\begin{document}

\title[A quantum no-reflection theorem and the speeding up of Grover's search algorithm]{A quantum no-reflection theorem and the speeding up of Grover's search algorithm}

\author{Karthikeyan S. Kumar and G.~S.~Paraoanu}
\address{Low Temperature Laboratory, Aalto University, P. O. Box 15100, FI-00076 AALTO, Finland, EU}
\ead{paraoanu@cc.hut.fi}

\begin{abstract}
We prove that it is impossible to built a universal quantum machine that produces reflections about an unknown state.
We then point out a connection between this result and the optimality of Grover's search algorithm: if such reflection machines were available, it would be possible to accelerate Grover's search algorithm to exponential speedups.
\end{abstract}
\maketitle

\section{Introduction}

A number of "impossibility" results exist in quantum mechanics, which illustrate the difficulty of replicating or extracting information from quantum objects. It is known for example that one cannot determine the wavefunction of a system by performing measurements on a single copy \cite{protective}. The rules of  quantum physics also prohibit the cloning of unknown states \cite{nocloning} as well as their reversible deleting \cite{nondeleting}. The general problem of characterizing quantitatively how well one can estimate a state is a fundamental issue in quantum mechanics and has attracted considerable interest \cite{stateestimation}.

In this paper we show that a related theorem holds -- it is impossible to build a machine which performs reflections about unknown states. The motivation for this result is similar to that of those above: we ask if certain operations, which are possible classically and are useful in standard computing, are also possible in quantum physics. For example, the no-cloning theorem is motivated by the need to store and replicate quantum information in the registers of a quantum processor. In our case, we would like to see wether it is possible to perform gates that realize reflections without having to measure (know) the state beforehand.

We then explore the connections between this theorem and the Grover's quantum search scheme \cite{grover}, in which such reflection operations appear naturally. Unlike other quantum algorithms, the improvement in this procedure with respect to classical search is only quadratic, and a mathematical proof showing that the Gover scheme is optimal exists \cite{zalka,nielsen}. We suggest here a connection between the generic problem of state estimation and the impossibility of universal reflection gates on one hand and the impossibility of accelerating  the Grover search algorithm on the other hand. More precisely, we show that, if restrictions on cloning and reflections were lifted, it would be possible to obtain a quantum search algorithm with exponential speed.

\section{Impossibility of reflection about unknown state}

\begin{figure}[t]
\begin{center}
  \includegraphics[width=7.5cm]{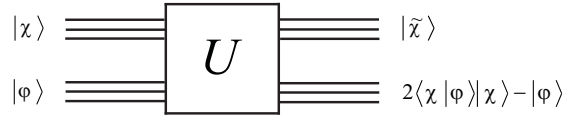}
\end{center}
\caption{Schematic of a quantum reflection gate.}
         \label{fig1}
\end{figure}

The generic operation of a reflection gate is shown in Fig. \ref{fig1}, implementing a unitary transformation U. In analogy with other two-qubit gates here the state $|\chi\rangle$ acts as a "control" and the state $|\varphi \rangle$  is the target to be modified. For example, the CNOT gate flips the state of the target qubit depending  on the state of the control qubit,
 $|c\rangle|t\rangle \stackrel{\rm CNOT}{=} |c\rangle|t\otimes c\rangle$. The resulting state of the target, $2\langle \chi|\varphi\rangle |\chi\rangle - |\varphi\rangle$, is a reflection of $|\varphi \rangle$ with respect to $|\chi\rangle$ \cite{nielsen}. The operator $2|\chi\rangle\langle\chi| -I$ is unitary for any state $|\chi\rangle$ is; this is expected also from its geometrical interpretation since reflections takes normalized states into normalized state. In the CNOT, the control qubit is left unchanged after the gate.  In our case, we do not impose anything on the output "control" wavefunction, which can be mapped into any wavefunction $|\tilde{\chi}\rangle$. Below we show that, even in this relaxed form, it is impossible to find a universal quantum circuit that would perform the desired reflection operation.

Consider the action of U on two sets of states $|\chi\rangle|\varphi\rangle$ and $|\chi '\rangle|\varphi '\rangle$,
\begin{eqnarray}
U|\chi\rangle|\varphi \rangle &=&  |\tilde{\chi}\rangle |[2\langle \chi|\varphi\rangle |\chi\rangle - |\varphi\rangle ] ,\\
U|\chi '\rangle|\varphi '\rangle &=&  |\tilde{\chi}'\rangle |[2\langle \chi '|\varphi '\rangle |\chi '\rangle - |\varphi '\rangle ] .
\end{eqnarray}
Taking the adjoint of the second expression and multiplying with the first one we obtain the condition
\begin{equation}
\langle \chi '|\chi\rangle - \langle \tilde{\chi} '|\tilde{\chi}\rangle =
2\left[ 2\langle\chi ' |\chi\rangle |\chi '\rangle\langle\chi| -  |\chi '\rangle\langle\chi '| - |\chi \rangle\langle \chi|\right] \langle\tilde{\chi}'|\tilde{\chi}\rangle .\label{nice}
\end{equation}
Let us now multiply Eq. (\ref{nice}) by $\langle\chi|$ on the left and by $|\chi \rangle$ on the right. We obtain
\begin{equation}
\langle \chi '|\chi \rangle = \left[2|\langle\chi '|\chi\rangle|^2-1\right]\langle\tilde{\chi}'|\tilde{\chi}\rangle . \label{ooo}
\end{equation}
Similarly, by multiplying to the left by $\langle\chi |$ and to the right by $|\chi '\rangle$ and using the fact that in general  $|\chi \rangle$ and $|\chi '\rangle$ need not be orthogonal, we get
\begin{equation}
\langle\chi '|\chi \rangle = \left[4|\langle\chi ' |\chi\rangle |^{2} - 3\right] \langle\tilde{\chi}'|\tilde{\chi}\rangle .\label{oooo}
\end{equation}
Combining Eq. (\ref{ooo}) and Eq. (\ref{oooo}) we get $|\langle \chi'|\chi\rangle|^2 = 1$ which forces the two states $|\chi \rangle$ and $|\chi '\rangle$ to be identical. Thus, we have shown that it is impossible to built a quantum circuit performing the desired operation shown in Fig. \ref{fig1}.

We now turn to the issue of accelerating Grover's search algorithm.

\section{The standard Grover search algorithm}

For completeness and to establish the notations, we now briefly review  the standard form of the Grover algorithm \cite{grover,nielsen}. For an unstructured data
base containing N elements, classically it takes ${\cal O}(N)$ queries to the database to find the solution to a search problem; in contrast, if quantum mechanics is used, only ${\cal O}(\sqrt{N})$ queries are required. The Grover algorithm requires $n=\ln N$ qubits which act as the index to the search elements and a register that act as the oracle workforce. The Grover operator is given by
\begin{equation}
G=(2|\psi\rangle\langle\psi|-I)O,
\end{equation}
where $|\psi\rangle$ is the initial uniform superposition state defined as
\begin{equation}
|\psi\rangle=\frac{1}{\sqrt{N}} \sum_{
0 \leq x \leq N-1
}|x\rangle ,
\end{equation}
 and $O$ is the oracle operator whose action is defined as
\begin{equation}
|x\rangle|q\rangle\rightarrow{O}|x\rangle |q\oplus f(x)\rangle.
\end{equation}
Here $x$ denotes the input to the oracle,
 $q$ is the oracle qubit and $f$ is defined as
$f(x) = 1$  if x is a solution to the search problem and $f(x)= 0$ otherwise.
The oracle qubit is chosen to be $(|0\rangle-|1\rangle )/\sqrt{2}$, and therefore the action of the oracle is
\begin{equation}
|x\rangle\frac{|0\rangle-|1\rangle}{\sqrt{
2}}\rightarrow{O}(-{}1)^{f(x)}|x\rangle\frac{|0\rangle-|1\rangle}{\sqrt{
2}}.
\end{equation}
Thus it can be seen that the oracle marks the solution to the search problem by flipping its phase. Its action can be represented by the simplified notation,
\begin{equation}
|x\rangle\rightarrow{O}(-1)^{f(x)}|x\rangle .
\end{equation}
as the oracle qubit is left unchanged during its action. The state $O|\psi\rangle$  is then acted upon by the unitary operator ($2|\psi\rangle\langle\psi| - I)$ to complete one Grover iteration.

The Grover operation can be visualized geometrically by defining a superposition of all non-solution states $|\alpha\rangle=\frac{1}{\sqrt{N-M}}\sum_x^{''}|x\rangle$, and the superposition of all states that are solutions to the search problem $|\beta\rangle=\frac{1}{\sqrt{M}}\sum_x^{'}|x\rangle$.
Here M denotes the number of solutions of the search problem. With these notations the initial state is
\begin{equation}
|\psi\rangle=\cos\frac{\theta}{2}|\alpha\rangle+\sin\frac{\theta}{2}|\beta\rangle ,
\end{equation}
where $\cos \theta /2=\sqrt{(N-M)/N}$ and $\sin\theta /2 =\sqrt{M/N}$.

Then one sees that the oracle causes a reflection about the $|\alpha\rangle$ axis whereas the $(2|\psi\rangle\langle\psi|-I)$  operating on $O|\psi\rangle$ causes it to get reflected about the $|\psi\rangle$ axis.
Thus each iteration rotates the state towards $|\beta\rangle$ by an angle $\theta$.
After $k$ iterations the resulting state is
\begin{equation}
G^k|\psi\rangle=\cos{\frac{(2k+1)\theta}{2}}|\alpha\rangle+\sin{\frac{(2k+1)\theta}{2}|}\beta\rangle .
\end{equation}
The total number of iterations $R$ is given by
$R\leq  \lceil \frac{\pi}{2\theta}  \rceil$,
and $\theta /2\geq\sin \theta /2=\sqrt{M/N}$.
Therefore $R\leq\lceil (\pi /4)\sqrt{N /M}\rceil$, or $
R={\cal O}(\sqrt{\frac{N}{M}})$.

\section{Exponential speedup: a modified search algorithm}

 Can we do in fact better? It is known that the answer is negative, at least for the case when the resources in terms of qubits are fixed \cite{zalka,nielsen}. But the proof uses rather abstract reasoning about unitary operations and one is left craving for a simpler argument. A debate on the issue of speeding up the search
 \cite{wrong} illustrates that the topic could be still confusing, and that a more intuitive approach to this problem is desirable.

What would be a simple way to accelerate the Grover algorithm beyond the standard
${\cal O}(\sqrt{N})$ number of searches? Intuitively, the situation can be described as follows: in the standard Grover algorithm, at the beginning of the step $k+1$ the quantum search machine has calculated anyway the state $|\psi_k\rangle$. Would we not be more efficient in using the information contained $|\psi_k\rangle$ if instead of performing reflections about the initial state $|\psi\rangle$
we perform reflections with respect to $|\psi_k\rangle$?

\begin{figure}
\includegraphics[width=7.5cm]{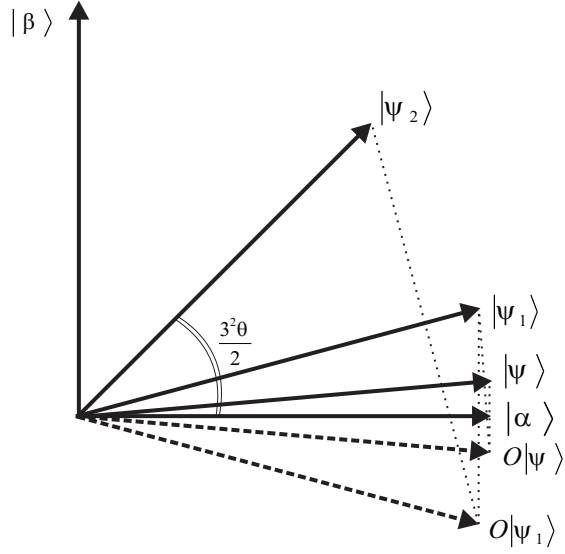}
\caption{Geometrical representation of the modified Grover algorithm}\label{fig2}
\end{figure}

Let us define an iteration-dependent operator such that $K_l |\psi_{l-1}\rangle=|\psi_l\rangle$ where $l$ is the iteration step as in Fig. \ref{fig2}. The operator $K_l$ is defined recursively as
\begin{equation}
K_{l+1}=(2|\psi_{l}\rangle\langle\psi_{l}|-I)O ,
\end{equation}
where initially $|\psi_{0}\rangle = |\psi\rangle $. The first iteration is identical to the standard Grover
$K_1=G=(2|\psi\rangle\langle\psi|-I)O$, but starting from the second iteration, the rotation of the state vector is accelerated. We show that after the step $l$, the state vector makes an angle $3^{l}\theta /2$ with the axis $|\alpha\rangle$. This follows immediately by induction. Indeed
\begin{equation}
|\psi_{1}\rangle = K_1|\psi\rangle \cos\frac{3\theta}{2}|\alpha\rangle+\sin\frac{3\theta}{2}|\beta\rangle,
\end{equation}
and, if $|\psi_{l}\rangle = \cos\frac{3^{l}\theta}{2}|\alpha\rangle +\sin\frac{3^{l}\theta}{2}|\beta\rangle$, then
\begin{eqnarray}
K_{l+1}|\psi_{l}\rangle &=&[2|\psi_{l}\rangle\langle\psi_l|-I]O|\psi_l\rangle
 \\
&=&\cos3^{l+1}\frac{\theta}{2}|\alpha\rangle+\sin3^{l+1}\frac{\theta}{2}|\beta\rangle .
\end{eqnarray}
At each iteration the state which had the initial phase of $\theta /2$ gets rotated
towards $|\beta\rangle$ by the angles $3\theta /2$, $9\theta /2$, $27\theta /2,...,$ up to $\pi /2$.
This is a geometric progression, thus one expects an exponential reduction in the number of steps needed. Now we want to find the number of iterations $R_{\rm mod}$ for our search algorithm; from the equation $\pi /2 = 3^{R_{\rm mod}}\theta /2$, we get
$R_{\rm mod}=\log_{3}\pi /\theta$, which together with $\sin
\theta /2\approx \theta /2=\sqrt{M/N}$
gives
\begin{equation}
R_{\rm mod}={\cal O}(\log_3\sqrt{\frac{N}{M}}) .
\end{equation}

What would one need to make this idea work in practice?  One first significant difficulty is the following: although in some sense the machine calculates at iteration $l$ anyway the state
$|\psi_l\rangle$, there is no place where the (complete) information about this state is stored in the machine so that it could be later recalled to construct the gates $K_l$. The reason for this is of course the no-cloning theorem.
One could still hope to extract enough information by performing a series of weak, nondisturbing measurements on the state $|\psi_l\rangle$. It is known however that this is not possible \cite{protective}. Using measurements in order to obtain classical information about a state and then moving it back in the quantum domain ({\it e.g.} to reconstruct the state or, as in our case, to built a gate) is not a good strategy. It has been shown that is more efficient to remain in the quantum domain and perform some type of quantum cloning \cite{measurements,galvaohardy}. Such quantum cloning algorithms have been developed recently \cite{duan,buzekhillery}, and it has been suggested that this technique can be used to increase the performance of certain quantum computation tasks \cite{galvaohardy}. As a method for distributing quantum information in a quantum processor, quantum cloning is known to be better than for example any type of measurement \cite{galvaohardy}. A discussion on the cloning techniques that can be used in this context is given in \ref{appendix}.

If we accept some loss of fidelity due to approximate cloning, or if we perform probabilistic cloning with unit fidelity, or if we just wish to understand what happens had the no-cloning theorem not been a stumbling block, then we can assume that after applying the oracle  to the state $|\psi_l\rangle$ we would  have available a copy of  $|\psi_l\rangle$ which can be used as a "control" $n$-qubit in order to implement a reflection about it. To implement $K_{l+1}$  we now need to generate a reflection of $O|\psi\rangle_{l}$ about $|\psi\rangle_{l}$ by using the clone of $|\psi\rangle_{l}$; this is how we would obtain the state
$2_{l}\langle\psi |\psi\rangle_{l+1} |\psi\rangle_{l} -|\psi\rangle_{l+1}$. This must be done at every step $l$, and since we do not know where the object is in the database, we also do not have any classical information about the states $|\psi\rangle_{l}$. This means that we would need to have a device that performs this operation for any input state - that is why we aim for a universal quantum circuit. At first sight, this looks like an innocuous operation, since we do not attempt to extract all the information from the state
$|\psi_l\rangle$ (only the projection $_{l}\langle\psi |\psi\rangle_{l+1}$).
 But, as we have seen, this cannot be constructed as a universal gate, according to the rules of quantum mechanics.

\section{Conclusions}
It is not possible to built a universal gate that would perform reflections about unknown quantum states. We link this result to the issue of optimality of Grover search. By an explicit scheme we show that, if the restrictions on cloning and reflections were not existent, it would be possible to accelerate the Grover algorithm to exponentially faster searches.

\section{Acknowledgments}
This work was supported by the Academy of Finland (Acad. Res. Fellowship 00857, and projects 129896, 118122, and 135135). The first author would like also to thank Prof. T.S. Mahesh (IISER Pune)
and Prof. R.R. Mishra (BITS-Pilani) for introducing him to the topic.

\appendix

\section{Probabilistic and approximate quantum cloning}
\label{appendix}

\begin{figure}[t]
\begin{center}
  \includegraphics[width=7cm]{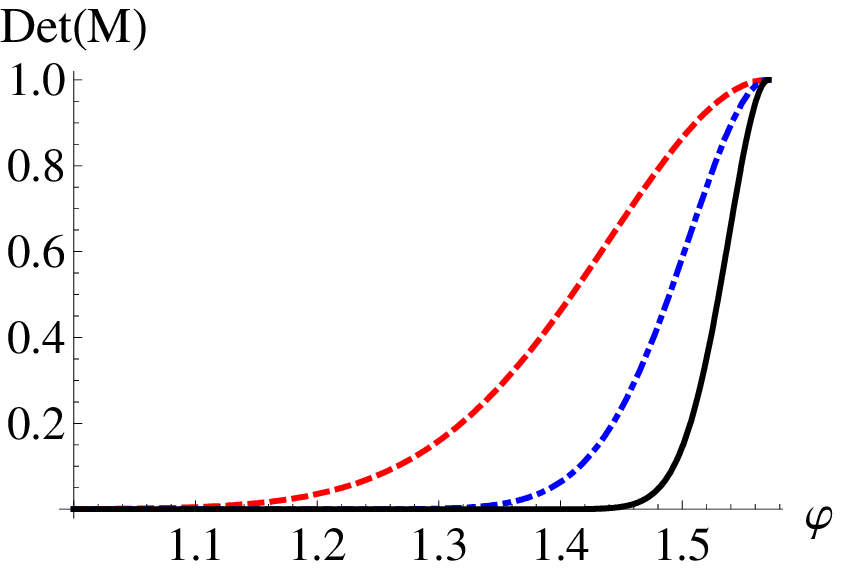}
\end{center}
\caption{The determinant of the matrix M for different-size registers $n=5$ (dashed), $n=7$ (dash-dotted), $n=9$ (continuous).}
         \label{fig3}
\end{figure}

Two types of quantum cloning algorithms have been discussed in the literature: probabilistic and deterministic cloning.
The first type uses postselection and produces perfect copies with a finite probability of success. The proof for the existence of these operations is constructive \cite{duan}. We discuss here the possibility of using such an algorithm for cloning the states $|\psi_l\rangle$ \cite{duan}, assuming for simplicity $M=1$ (one item to search).  The general structure of these states (both in the standard and the modified version of the Grover algorithm) is
\begin{equation}
|\psi\rangle = \sin \varphi |\beta\rangle + \cos \varphi |\alpha\rangle , \label{gen}
\end{equation}
where $|\beta\rangle$ is the state we are searching for. Since we do not know in general where the item is, it means that we do not know, out of all possible vectors $|x\rangle$, which one is the vector $|\beta\rangle$; however, if we establish that the states $|\psi\rangle$ form a linearly independent set for all the possible choices of $|\beta\rangle$ from the basis  $\{|x\rangle\}$, then one  can probabilistically clone these states. To check for the linear independence of these state, we calculate the determinant of
a matrix $M$ with $\sin \varphi$ on the diagonal and $(1 /\sqrt{N-1}) \cos \varphi$ as the other entries (see Fig. \ref{fig3}). By numerical testing, we found that for relatively large registers this determinant is positive but close to zero for most of the initial part of the calculations, and it becomes nonzero only in the last part. This is intuitively expected: indeed, one can notice that the starting state in the Grover algorithm is the same (no matter where the item is). When iterating, the states corresponding to different solutions of the search begin to differentiate with respect to each other due to amplification of the amplitude probability of the solution. The value 1 at $\theta = \pi /2$ corresponds to having found the solution to the problem, which is one of the states $|x\rangle$; these states are orthogonal and Det(M) = 1 in this case. Although probabilistic cloning of these states is in principle possible, we notice that the smallness of the determinant for most of the angles means that probabilistic cloning can work effectively only in the latest iterations of the search algorithm.

 The second type of quantum cloning is deterministic and universal, yielding approximate copies of the original with  a certain fidelity \cite{buzekhillery}. In our case, we would need to clone a quantum register. In the situation of creating two clones, the so-called scaling factor $s$ \cite{buzekhillery} is
\begin{equation}
s= \frac{N+2}{2(N+1)} ,
\end{equation}
and the fidelity of the two copies with respect to the original is
\begin{equation}
F= \frac{1-s}{N}+s = \frac{N+3}{2(N+1)} .
\end{equation}
An interesting observation is that even in the limit $N\rightarrow \infty$ the fidelity is finite, $F=1/2$. For a large register, the use of this technique even over a few steps would introduce considerable errors.

\end{document}